\documentclass[useAMS,usenatbib]{mn2e}

\usepackage{epsf}

\def\kms{$\rm km\, s^{-1}$}
\def\cm3{$\rm cm^{-3}$}

\def\n0{$\rm n_{0}$}
\def\B0{$\rm B_{0}$}

\def\T r{$\rm T_{ r}$}
\def\T as{$\rm T_{ as}$}

\def\um{$\rm \mu m$}

\title{Morphology of the coronal line region  in active galactic nuclei}
\author[M. Almudena Prieto]{M. Almudena Prieto$^{1}$, Olivier Marco$^{2}$,Jack Gallimore$^{3}$\\
$^{1}$Max-Planck fuer Astronomie,  Heidelberg, Germany; prieto@mpia.de \\
$^{2}$ESO Paranal, Alonso de Cordova 3107, Vitacura, Santiago, Chile \\
$^{3}$Department of Physics, Bucknell University, Lewisburg, PA
  17837, USA
}

\begin{document}

\maketitle

\begin{abstract}
We present new images\footnote{Observations done under 
ESO/VLT  programs  70.B-0409 and 74.B-0404} of the
coronal line region, as traced by [Si VII] 2.48~\um, in some of the
nearest Seyfert 2 galaxies.  In each of these galaxies, the coronal
line emission comprises a bright, compact central source and extended
emission showing broad alignment along a particular direction, usually
coinciding with that defined by the radio emission or the extended
narrow line region.  The full extent
of the coronal line emission ranges from few tens of pc to $\sim$ 150
pc radius from the nucleus and is a factor $\sim 10$ smaller than that
seen in the extended, lower-ionization gas. With a spatial resolution
of 10 pc or better, the coronal region shows diffuse and filamentary
structure in all cases, and it is difficult to see whether it 
breaks down into discrete blobs as those seen in
lower-ionization-lines- or radio- images of comparable resolution. The
extent of the coronal line emission is larger than would be predicted
by photoionization models, which argues for additional in-situ gas
excitation, the most plausible energy source being shock-excitation.
\end{abstract}

%%\keywords{galaxies:nuclei -- galaxies:Seyfert -- infrared:galaxies}


\section{Introduction}
Coronal lines are collisionally excited forbidden transitions within
low-lying levels of highly ionized species (IP $>$ 100 eV). As such,
these lines  form in extreme energetic environments and thus
are unique tracers of AGN activity; they are not seen in starburst
galaxies.  Coronal lines appear from X-rays to IR and are common in
Seyfert galaxies regardless of their type (Penston et al.  1984;
Marconi et al. 1994; Prieto \& Viegas 2000). 
%In the optical, the most
%important coronal lines are from the Fe series: [Fe VII] to [FeXIV]. 
The strongest ones  are seen  in the IR; in the
near-IR they can even dominate the line spectrum (Reunanen et al. 2003).

Owing to the high ionization potential, these lines 
%should be produced
%very close to the active nucleus where the density of high energy
%photons is large, and furthermore the electron density is also
%expected to be relatively large. The high critical density of these
%lines, $10^7 -10^{10}  cm^{-3}$, guaranty their formation in 
%such environment
%Based on photoionization
%considerations, the coronal line region (CLR) is 
are expected to be
limited to few tens to hundred of parsec around the active nucleus.
% At the distance of even the nearest Seyfert galaxies, the
%CLR would be spatially unresolved in seeing-limited
%observations.
%Among Seyfert galaxies, only NGC~1068 has been imaged with the HST at
%the position of the IR coronal line [SiVI] 1.963~\um; the extent of
%the emission ranges up to 300 pc from the center (Thompson et
%al. 2001). The difficulty in interpreting this observation, however,
%is that [SiVI] 1.963~\um\ is heavily blended with H$_{2}$
%1.957~\um\, which we know from IR spectroscopy is both strong and
%extended up to several hundred parsec radius from the nucleus
%(Rodriguez-Ardila et al. 2004).  In contrast, the nearer Circinus
%Galaxy shows extended [Si VI] 1.963~\um\ up to 30 pc from the nucleus
%(Maiolino et al. 1998). This is a robust result as in this case [Si
%VI] 1.963~\um\ image was extracted from imaging spectroscopy data
%which permits separation from the 1.957~\um\ line.
On the basis of spectroscopic observations, 
Rodriguez-Ardila et al. (2004, 2005) 
%have conducted a high spectral
%and spatial spectroscopic resolution survey of optical and IR coronal
%lines of the nearest AGN accessible from the ESO observatories.
%Although those observations are seeing-limited,
% ($<~ 1$ arcsec in the optical; $\sim$ 0.5 arcsec in the IR),
%the relatively high spectral resolution, $\sim 100$~\kms, and
%sensitivity of the data 
unambiguously established the size of the coronal line region (CLR)
in NGC~1068 and the Circinus Galaxy, using  the coronal
lines [SiVII] 2.48~\um\,, [SiVI] 1.98~\um\~,, [Fe\,{\sc vii}] 6087\AA,
[Fe\,{\sc x}] 6374 \AA\ and [Fe\,{\sc xi}] 7892 \AA\. They find these 
lines  extending up
to 20 to 80 pc from the nucleus, depending on ionization potential.
Given those sizes, we started 
an adaptive-optics-assisted imaging program with the 
ESO/VLT aimed at
revealing the detailed morphology of the CLR in some of nearest
Seyfert galaxies. We use as a tracer the isolated IR line [Si VII]
2.48~\um\  (IP=205.08 eV). 
% A sample of the nearest Seyfert galaxies
%known to present strong [Si VII] emission based on previous
%spectroscopic studies (Reunanen et al. 2002, 2003) was selected.  
This letter presents the resulting narrow-band images of the [Si VII]
emission line, which
%from the highest quality VLT/NACO observations that we
%have gathered to date. These images 
reveal for the first time the detailed morphology of
the CLR, and with suitable resolution for comparison with radio and 
optical-
%images (particularly narrow-band images which trace the
lower-ionization-gas images. The morphology of the CLR is sampled
with a spatial resolutions almost a factor 5 better than any
previously obtained, corresponding to scales $\sim <$10 pc.  The
galaxies presented are all Seyfert type 2: Circinus, NGC~1068,
ESO~428-G1 and NGC~3081. Ideally, we had liked to image type 1
objects, but, in the Southern Hemisphere, there are as yet no known,
suitable type~1 sources at sufficiently low redshift to guarantee the
inclusion of [Si VII] 2.48~\um\ entirely in the filter pass-band.

\section{Observations, image registration and astrometry}

Observations  were done  with the adaptive-optics assisted IR
camera NACO at the ESO/VLT.  Two narrow band filters, one centered on
the coronal [SiVII] 2.48~$\mu m$ line and  an adjacent band centered on
2.42~$\mu m$ line-free continuum, were used.  The image scale was
0.027 arcsec pixel$^{-1}$ in all cases, 0.013  arcsec  pixel$^{-1}$
in NGC\,1068.  
Integration times
were chosen to keep the counts within the linearity range: $\sim 20$
minutes per filter and source.  
For each filter, the photometry was
calibrated against standard stars observed after each science target.
These stars were further used as PSF when needed and for deriving  a 
correction   factor that
normalizes    both narrow-band filters to provide
  equal number of counts 
per a given flux. In deriving this factor it is assumed that 
the continuum level
in the stars  is the same in both filters and not emission lines are
not present.
%The redshifts of the selected galaxies are low enough to guarantee
%that [SiVII] 2.48~$\mu m$ line falls inside the filter width.  
The wavefront sensor of the adaptive optics system followed the optical
nucleus of the galaxies to determine seeing corrections. 

The achieved spatial resolution was estimated from stars available in
the field of the galaxies when possible; this was not possible in 
NGC 3081 and NGC 1068 (cf. Table 1). 
The resolutions were  comparable in both filters
within the reported errors in Table 1. 
Continuum-free [SiVII]2.48~\um\ line images are shown 
in Figs. 1 and 2 for each
galaxy. These were produced after applying the normalization factor
derived from the standers stars. 
The total integrated coronal line emission derived from these
images is listed in Table 2.  For comparison, [SiVII] 2.48~\um\ fluxes
derived from long-slit spectroscopy are also provided.

Also in these figures, images 
 with the  2.48~\um\~ filter of the   standard stars -also used as PSF's control- are
 shown. The images provide a rough assessment of 
the image quality/resolution achieved in the 
 science frames. For the case of Circinus and  ESO 428-G014, a more
 accurate evaluation  is possible   from  the images of 
a field star. One of these field star is shown  in both  filters   
in Figs. 1e  and 2b respectively..  To get an assesment of the
 image quality at the lowest signal levels, the images of the field stars in particular are  
 normalized to  the galaxy peak at the corresponding filter. These 
are much fainter than the galaxy nucleus, thus, the 
 star peak  is a mere $\sim$5\% of the galaxy peak.

%\begin{deluxetable*}{lccccccc}
\begin{table*}
\centering
%\begin{minipage}{140mm}
\caption{Galaxies scale and achieved NACO angular resolution. $*$: in NGC 1068,  is given the
size of the nucleus as K-band interferometry  sets an upper limit for the 
core of 5 mas (Weigelt et al. 2004);  in NGC 3081, the size of a PSF  star  
taken after the science frames is given}
%\tablehead{
\begin{tabular}{cccccccc}
 \hline 
AGN & Seyfert & 1 arcsec & Stars & FWHM & FWHM & Size of nucleus \\
 &type &in pc & in field & arcsec &   pc &   FWHM arcsec \\
%\startdata
 \hline
Circinus	& 2 & 19   & 2	& 0.19$\pm$0.02	  & 3.6 &  0.27 \\
NGC 1068        & 2 & 70   & 0  & 0.097$^*$        & 6.8 & $<$0.097 \\
ESO\,428-G014   & 2 & 158  & 3  & 0.084$\pm$0.006 & 13  & 0.15$\pm$0.01 \\
NGC\,3081       & 2 & 157  & 0  & 0.095$^*$       & 14 & $<$0.32 \\
 \hline
%\enddata
 \end{tabular}

%\end{minipage}
\end{table*}

\begin{table*}
%\tablewidth{0pt}
\centering
\caption{Size and photometry of the 2.48 $\mu m$ coronal line region.  $*$: from a 1'' x 1.4'' aperture  \label{flxTb}}
 \begin{tabular}{ccccc}
\hline
{AGN} &  {Radius from nucleus} &  {Flux NACO} &   {Flux long-slit} &   {Reference long-slit} \\
&pc &\multicolumn{2}{c}{in units of $10^{-14}$~erg~s$^{-1}$ cm$^{-2}$} \\

%\startdata
\hline
Circinus  & 30   & 20 & 16 &  Oliva et al. 1994 \\
NGC 1068 & 70  &  72 & 47$^*$ & Reunanen et al. 2003 \\
ESO\,428-G014 & 120 - 160  & 2.0	& 0.7$^*$ & Reunanen et al. 2003 \\
NGC\,3081 & 120	& 0.8	& 0.8$^*$ & Reunanen et al. 2003 \\
%\enddata
\hline
  \end{tabular}
%\footnote{*}{From a 1'' x 1.4'' aperture}
\end{table*}

Radio and HST images were used, where available, to establish an
astrometric reference frame for the CLR in each of the galaxies.
%Accurate registration of the images was done on the basis of available
%common stars in the NACO field. However, as detailed below, this was
%not always possible, particularly for the radio images or pure
%emission line images.  
For NGC~1068 (Fig 1a, b, \& c), the registration of radio,
optical HST and adaptive-optics IR images by Marco et
al. (1997;  accuracy $\sim$ 0.05'') was adopted.  The comparison of the [SiVII]
2.48~\um\ line image with the HST [OIII] 5007 \AA\ followed by 
 assuming the peak emission in Marco's et
al. K-band image to coincide with that in the NACO 2.42~\um\ continuum
image.  The comparison with the MERLIN 5~GHz image of Gallimore et
al. (2004) was done assuming that the nuclear radio source 'S1' and 
the peak emission in the NACO 2.42~\um\ image are coinciding.

In Circinus (Fig. 1d, e \& f), the registration
of NACO and HST/$H\alpha$ images was done on the basis of  3--4
stars or unresolved star clusters available in all  fields.
 That provides an accurate registration better
than 1 pixel (see Prieto et al. 2004).  No radio image of comparable
resolution is available for this galaxy.

For ESO\,428-G014 (Fig. 2a, b, \& c), NACO images were registered on
the basis of 3 available stars in the field. Further registration with
a VLA 2~cm image (beam 0.2''; Falcke, Wilson \& Simpson 1998) was made
on the assumption that the continuum peak at 2.42~$\mu m$ coincides with
that of the  VLA core.  We adopted the astrometry provided by
Falcke et al. (uncertainty $\sim$ 0.3") 
who performed the registration of the 2~cm and the
HST/H$\alpha$ images, and plotted the HST/H$\alpha$ atop the NACO
coronal line image following that astrometry. 

NGC~3081 (Fig. 2d, e, \& f) has no stars in the field.  In this case
NACO 2.42~\um\ and 2.44~\um\ images, and an additional NACO deep
Ks-band image, were registered using the fact that the NACO adaptive
optics system always centers the images at the same position of the
detector within 1 pixel (0.027'').  The registration with a HST/WFPC2
image at 7910\AA\ (F791W), employed as a reference the outer isophote
of the Ks-band image which show very similar morphology to that seen
in the HST 7910\AA\ image.  Further comparison with an HST PC2
$H\alpha$ image relied on the astrometry by Ferruit et al. (2000).
The registration with an HST/FOC UV image at 2100\AA\ (F210M) was
based on the assumption that the UV nucleus and the continuum peak emission at 2.42 2.42~\um\  coincides.
The radio images available for this galaxy have a beam
resolution $>0.5''$ (Nagar et al. 1999), which includes all the
detected coronal extended emission, and are therefore not used in this
work.

\section{The size and morphology of the coronal line region}

In the four galaxies, the CLR resolves into a bright
nucleus and extended emission along a preferred position angle, which
usually coincides with that of the extended lower-ionization gas. The
size of the CLR is  a factor 3 to 10 smaller than the extended
narrow line region (NLR). The maximum radius (Table 2) varies from 30
pc in Circinus to 70 pc in NGC 1068, to $\sim >$ 120 pc in NGC 3081
and ESO~428-G014. The emission in all cases is diffuse or filamentary,
and it is difficult to determine whether it further breaks
down into compact knots or blobs such as those found in H$\alpha$, [OIII]
5007\AA\ or radio images even though the resolutions are comparable.

In Circinus, [SiVII]2.48~\um\ emission extends across the nucleus and
aligns with the orientation of its one-sided ionization cone, seen in 
H$\alpha$, or in [OIII] 5007 \AA. In these lines, the counter-cone is not seen (Wilson et
al. 2002), but in [SiVII], presumably owing to the reduced extinction,
extended diffuse emission is detected at the counter-cone position
(Fig. 1f; Prieto et al. 2004). 
%[SiVII] extends $\sim$30 pc to the
%northwest along the cone direction, 
%and $\sim$25pc to the southeast of the
%nucleus, in the counter-cone direction. There is extended 
This has been  further confirmed with
VLT/ISAAC  spectroscopy
which shows both [SiVII]2.48~\um\ and [SiVI] 1.96~\um\ extending up to
30 pc radius from the nucleus (Rodriguez-Ardila et al. 2004).
In the coronal line image, the  North-West  emission
is  defining an   opening cone angle larger than that in  $H\alpha$.
The  morphology of [SiVII] in this region is suggestive of
 the coronal emission tracing the
walls of the ionization cone (see fig. 1f).

In ESO~428-G014, the coronal emission is remarkably aligned with
the radio-jet (Fig. 2c). The 2~cm emission is stronger in the
northwest direction, and [SiVII] is stronger in that direction too.
H$\alpha$ emission is also collimated along the radio structure, but
the emission spreads farther from the projected collimation axis and
extends out to a much larger radius from the nucleus than the coronal
or radio emission (Fig. 2b).  Both H$\alpha$ and the 2 cm emission
resolve into several blobs but the coronal emission is more diffuse.

In NGC 3081, the coronal emission resolves into a compact nuclear
region and a detached faint blob at $\sim$120 pc north of it.  The HST
[OIII] 5007 and H$\alpha$ images  show rather collimated structure
extending across the nucleus along the north-south direction over
$\sim$ 300 pc radius (Ferruit et al. 2000).  
%The optical morphology
%further resolves into several concentric shell-like regions.
%(this morphology shows clearly in Fig. 11 of Ferruit et al.).  
Besides the nucleus, the second brightest region in those lines 
coincides with the detached 
[Si VII] emission blob (Fig. 2d).  At this same position, we also find
UV emission in a  HST/FOC image at 2100 \AA.

NGC~1068 shows the strongest [Si VII] 2.48~\um\ emission among the four
galaxies, a factor three larger than in Circinus, and the only case
where the nuclear  emission shows detailed structure. At 
$\sim 7 ~pc$ radius from the radio core S1, [Si VII] emission
divides in three bright blobs.
The position of  S1 falls in between the blobs. The southern blob looks
 like a concentric shell. The northern blob coincides with the
central [OIII] peak emission at the vortex of the ionization cone; the
other  two  blobs are not associated with a particular
enhancement in [OIII] or radio emission (Fig. 1b \& c).  [Si
VII] depression at the position of S1 may indicate a very high
ionization level at already 7 pc radius (our resolution)  from the center; the
interior region might instead be filled with much higher ionization-level
gas, e.g.  [FeX], [Si IX] and higher.  This central structure, $\sim$14~ pc 
radius in total, is  surrounded in all directions by much lower
surface brightness  gas, extending up to at least 70 pc
radius.  The presence of this diffuse region is
confirmed by VLT/ISAAC spectra along the north-south direction, which
reveal [SiVI]1.96~\um\ and [Si VII]2.48~\um\ extending at both sides of
the nucleus up to comparable radii (Rodriguez-Ardila et al. 2004, 2005).  This diffuse 
emission shows slight enhancement at both sides of the 5 GHz jet, but
there otherwise appears no direct correspondence between the CLR and
radio morphology.

\section{Discussion}

ESO~438-G014 and NGC~3081 show the largest and best collimated [SiVII]
emission, up to 150 pc radius from the nucleus. To reach those
distances by nuclear photoionization alone would require rather low
electron densities or a very strong (collimated) radiation field.
Density measurements in the CLR are scarce: Moorwood et al. (1996)
estimate a density $n_e \sim 5000~cm^{-3}$ in Circinus on the basis of
[NeV]~14.3~\um\ /24.3~\um; Erkens et al. (1997) derive $n_e < 10{^7}~
cm^{-3}$ in several Seyfert 1 galaxies, on the basis of several
optical [FeVII] ratios. This result may be uncertain because the
optical [Fe VII] are weak and heavily blended.  Taking  $n_e \sim 10^4
cm^{-3}$ as a reference value, it  results in an ionization parameter U
$<\sim 10^{-3}$ at 150~pc from the nucleus, which is far too low to produce strong [SiVII]
emission (see e.g Ferguson et al. 1997; Rodriguez-Ardila et
al. 2005). 

We argue that, in addition to photoionization, shocks must
contribute to the coronal emission.  This proposal is primarily
motivated by a parallel spectroscopic study of the kinematics 
%excitation and  size 
of the CLR gas of several Seyfert galaxies
(Rodriguez-Ardila et al. 2005), which reveals 
%shows  that photoionization
%is not sufficient to explain the size and the excitation of the
%coronal gas; yet 
coronal line profiles with  velocities  500 \kms $< v <$ 2000 \kms .
% -- as indicated by the coronal line profiles -- can provide the additional
%energy input needed.  
Here we assess the proposal in a qualitative
manner, by looking for evidence for shocks from the morphology of the
gas emission.

In ESO~428-G014, the remarkable alignment between [Si VII] and the radio emission
is a strong indication of the interaction
of the radio jet with the ISM. There is  spectroscopic
evidence of a highly turbulent ISM in this object: asymmetric
emission line profiles at each side of the nucleus indicate gas velocities
of up to 1400 \kms (Wilson \& Baldwin 1989).  Shocks with those
velocities heat the gas to temperatures of $>\sim 10^7 K$, which will
locally produce bremsstrahlung continuum in the UV -- soft X-rays
(Contini et al. 2004) necessary to produce coronal lines.  [Si VII]
2.48~\um\ with IP = 205.08 eV eV will certainly be enhanced in this process.

The concentric shell-like structure seen in NGC 3081 in [OIII] 5007
\AA\ and H$\alpha$ (Ferruit et al. 2000) 
is even more suggestive of propagating shock
fronts.  From the [OIII]/H$\alpha$ map by Ferruit et al., the
excitation level at the position of the [Si VII] northern blob is similar to
that of the nucleus, which points to similar ionization parameter
despite the increasing distance from the nucleus. The cloud density might then 
decrease with distance to balance the
ionization parameter, but this would demand a strong radiation field
to keep line emission efficient. Alternatively, a local source of
excitation is needed. The presence of cospatial UV
continuum, possibly locally generated bremsstrahlung, and [Si VII]
line emission circumstantially supports the shock-excitation proposal.

In the case of Circinus and NGC~1068, the direct evidence for shocks
from the [Si VII] images is less obvious. In NGC~1068, the orientation
of the three blob nuclear structure does not show an obvious
correspondence with the radio-jet; it may still be  possible  we missed 
the high velocity coronal gas component measured in NGC 1068   
in our narrow band filter.
 In Circinus, there are not radio
maps of sufficient resolution for a meaningful comparison.  However, 
both galaxies present high velocity nuclear outflows, which are
inferred from the asymmetric and blueshifthed profiles measured in the
[OIII] 5007 gas in the case of Circinus (Veilleux \& Bland-Hawthorn
1997), and in the Fe and Si coronal lines in both. In the latter,
velocities of $\sim$500 \kms in Circinus and $\sim$ 2000 \kms in NGC
1068  are inferred from the coronal profiles (Rodriguez-Ardila et
al. 2004, 2005). 

An immediate prediction for the presence of shocks is the production
of free-free emission, with a maximum in the UV- -- X-ray, 
from the shock-heated
gas. We make here a first order assessment of this contribution using
results from photoionization - shocks composite models by Contini et
al. (2004), and compare it with the observed soft X-rays.
  For each galaxy, we derive the 1 keV emission due to
free-free from models computed for a nuclear ionizing flux, $F_h = 10^{13}
photons~cm^{-2} s^{-1} eV^{-1}$, pre-shock density $n_o=300 cm^{-3}$ and shock velocity closer
 to the gas velocities measured in these galaxies (we use
figure A3 in Contini et al.). The selection of this high-
ionizing-flux  value
has a relative low impact does  on   the 1 keV emission estimate
 as the    bremsstrahlung emission from this flux  drops sharply shortwards the Lyman 
limit;  the results are  more depending on the strength of the 
post-shock bremsstrahlung  component, this being  mainly dominated 
by the shock  velocity and peaks in the soft X-rays (see fig. A3 in Contini et al. 
for illustrative examples). Regarding selection of densities,   
 pre-shock densities of a few hundred  $ cm^{-3}$  
actually imply densities downstream (from where 
 shock-excited lines are  emitted) a factor of 10 - 100 higher, 
the higher for higher velocities, 
and thus  whitin the range of those estimated from  coronal line measureemnts (see above).

Having selected the model parameters, we further assume that the estimated   1 keV emission 
comes from a region with size that of the observed [Si VII] emission.
Under those premises, the results are as follows.  For NGC 1068,
assuming the free-free emission extending uniformly over a $\pi
\times (70 pc)^2 cm^{-2}$ region (cf. Table 1), and models for shock
velocities of 900 \kms, the inferred X-ray flux is larger by a factor
of 20 compared with the nuclear 1 keV Chandra flux derived by Young et
al. (2001).  One could in principle account for this difference by
assuming a volume filling factor of 5-10\%, which in turn would
account for the fact that free-free emission should mostly be produced
locally at the fronts shock.
  
In the case of Circinus, following the same procedure, we assume a
free-free emission size of $\pi \times (30 pc)^2 cm^{-2}$ (cf. Table
1), and models of shock velocities of 500 \kms (see above). In this
case, the inferred X-ray flux is lower than the 1 keV BeppoSAX flux,
as estimated in Prieto et al. (2004),  by an order of magnitude.  For
the remaining two galaxies, we assume respective free-free emission
areas (cf. Table 1) of 300 pc x 50 pc for ESO 428-G014 --  the width of [Si VII] is  $\sim 50 pc$ in the direction
perpendicular to the jet -- and $2 \times (\pi \times (14~pc)^2
cm^{-2})$ for NGC 3081 -- in this case, free-free emission is assumed to 
come from
the nucleus and the detached [Si VII] region North of it only. Taking
the models for shocks velocities of 900 \kms, the inferred X-ray
fluxes, when compared with 1 KeV fluxes estimated from BeppoSAX data
analysis by Maiolino et al. (1998), are of the same order for ESO
428-G014 and about an order of magnitude less in NGC 3081.

The  above results are clearly  dominated by the assumed size of the
free-free emission region, which is unknown.  The only purpose of this
exercise is to show that under reasonable assumptions of shock
velocities, as derived from the line profiles, 
the   free-free emission  generated by these shocks  in the 
 X-ray  could be accommodated within the observed soft
X-ray fluxes.

We thank Heino Falcke who provided us with the 2 cm radio image 
of ESO 428-G014, and Marcella Contini for a thorough review of the manuscript.

\clearpage

\begin{figure*}
\begin{center}
%\includegraphics[scale=0.85]{Figs1.eps}
%\includegraphycs[scale=0.85]{f1.eps}
\vspace{-0.5cm} 
\epsfxsize=20cm
 \epsfxsize=20cm
\epsfbox{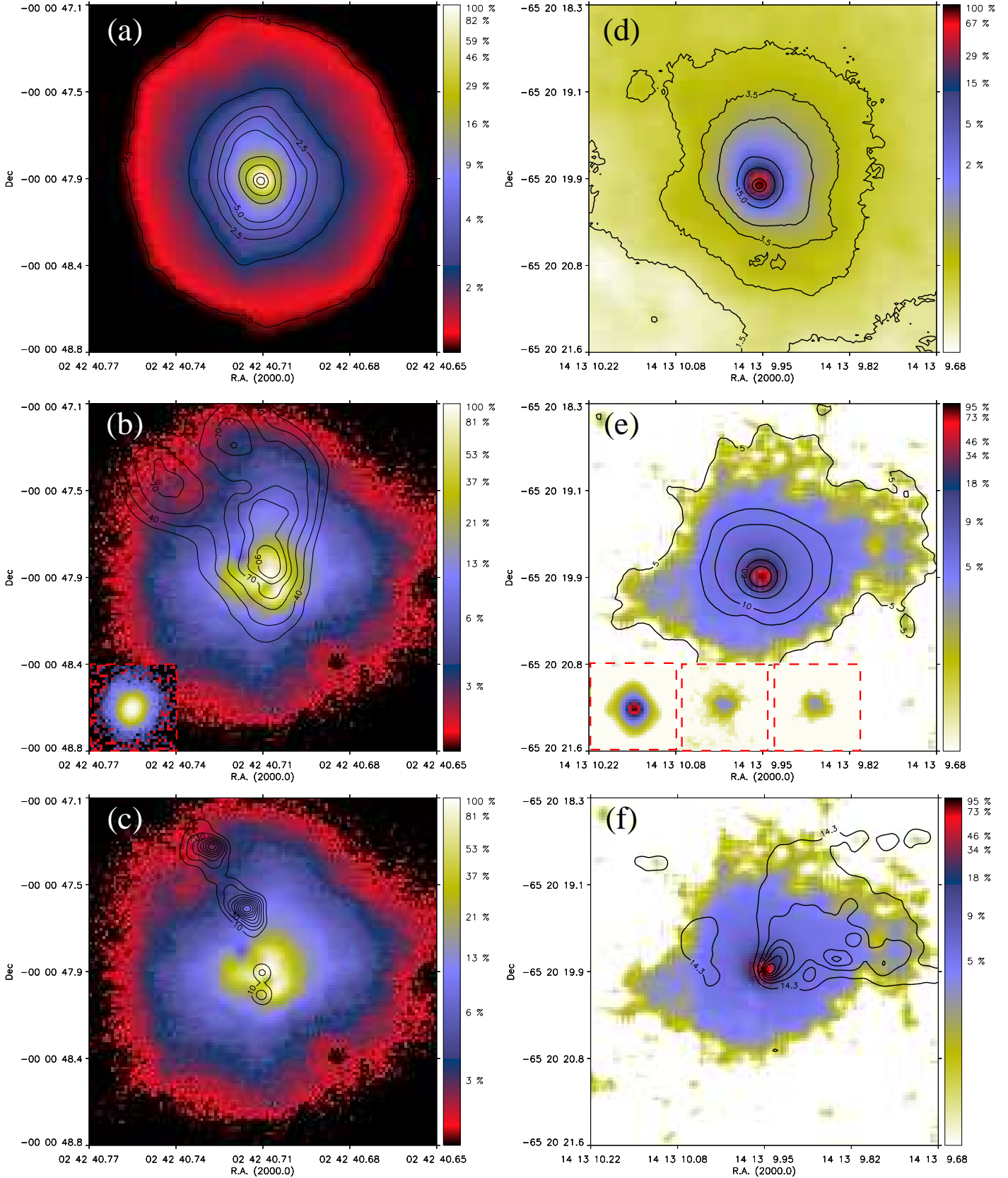}
\end{center}
\vspace{-0.4cm} 
\caption{NACO images (log scale) overlaid with contours (unless
  indicated numbers in contours 
and in the color bar are in percent of the  maximum peak 
for both, galaxies and stars): 
a) 2.42~\um\ continuum.
b) [SiVII]2.48~\um\ and HST [OIII]5007\AA\ (contours). Inset:
standard star in  2.48~\um\ filter.
c) [SiVII]2.48~\um\ and MERLIN 5 GHz (Gallimore et al. 2004).
d)  2.42~\um\ continuum.
e)  [SiVII]2.48~\um\ line. Insets: left,
standard star in  2.48~\um\ filter; center and right:
  field star at 9.1 arcsec from nucleus and PA= -30 deg, in  2.42~\um\ and 2.48~\um\ filters
  respectively,  maximum normalized to  the galaxy peak at that filter.
f)  [SiVII]2.48~\um\ and  HST H$\alpha$ (contours)
}
\label{}
\end{figure*}
\clearpage

\begin{figure*}
\begin{center}
%\includegraphics[scale=0.85]{Figs2.eps}
%\includegraphycs{f2.eps}
\vspace{-0.6cm} 
\epsfxsize=20cm
 \epsfxsize=20cm
\epsfbox{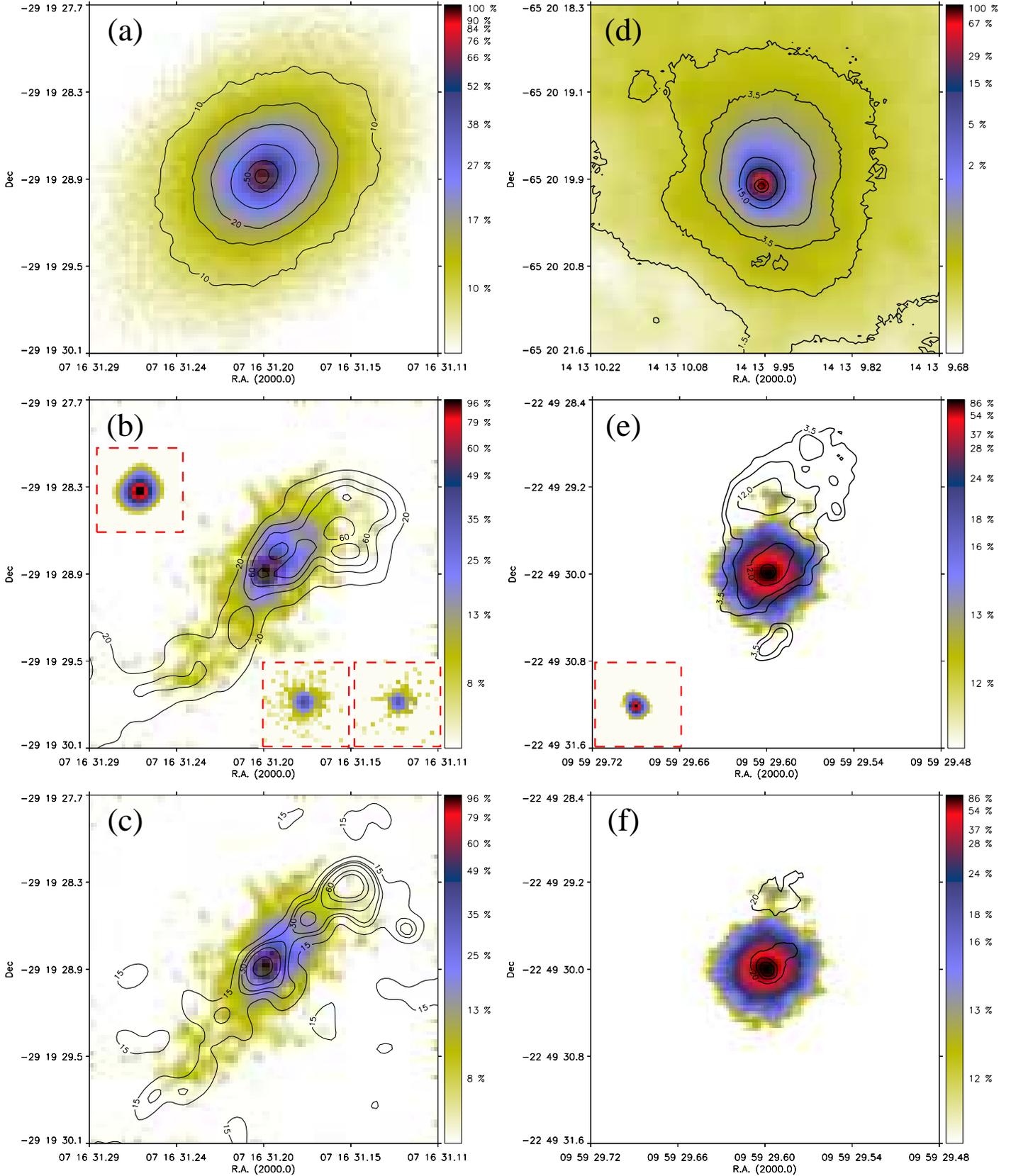}
\end{center}
\caption{NACO images (color code   and contours numbers as in Fig. 1):
a) 2.42~\um\ continuum
b)  [SiVII]2.48~\um\ and  HST  H$\alpha$ (contours). Insets: top,
standard star in  2.48~\um\ filter; bottom center and right:
field star at 11 arcsec from nucleus and PA = 7 deg, in  2.42~\um\ and 2.48~\um\ filters
  respectively, normalized to  the galaxy peak at that filter.
c)  [SiVII]2.48~\um\ and  2\,cm VLA image (Falcke et al. 1998, contours)
d) 2.42\um\ continuum
e)  [SiVII]2.48~\um\ and HST H$\alpha$ (contours). Inset:
standard star in  2.48~\um\ filter.
f)  [SiVII]2.48~\um\ and HST UV~2100\AA\ (contours)
}
\label{}
\end{figure*}

\end{document}